\documentclass[prl, 10pt,letterpaper,superscriptaddress,floatfix,nofootinbib,notitlepage,twocolumn]{revtex4-1}

\usepackage{textcomp}
\usepackage{amsmath}
\usepackage{amsfonts}
\usepackage{siunitx}
\usepackage{amssymb}
\usepackage{graphicx}
\usepackage{siunitx}
\usepackage{caption}
\usepackage{setspace}
\usepackage{xcolor}
\usepackage{graphicx}
\usepackage{lipsum} 
\usepackage{afterpage}
\usepackage{caption}
\usepackage[resetlabels,labeled]{multibib}
\newcites{BC}{BC Readings}

\setcitestyle{super}
\usepackage[colorlinks, citecolor={blue}, linkcolor={black}]{hyperref}

\DeclareMathOperator*{\argmax}{arg\,max}
\DeclareMathOperator*{\argmin}{arg\,min}

\begin{document}
\raggedbottom

\title{On-Chip Multidimensional Dynamic Control of Twisted Moiré Photonic Crystal for Smart Sensing and Imaging}

\author{Haoning Tang $^\bot$}
\email{hat431@g.harvard.edu}
\affiliation{School of Engineering and Applied Sciences, Harvard University, Cambridge, MA 02138, USA}

\author{Beicheng Lou $^\bot$}
\affiliation{Department of Applied Physics and Ginzton Laboratory, Stanford University, Stanford, CA 94305, USA}

\author{Fan Du}
\author{Guangqi Gao}
\author{Mingjie Zhang}
\author{Xueqi Ni}
\author{Evelyn Hu}
\affiliation{School of Engineering and Applied Sciences, Harvard University, Cambridge, MA 02138, USA}

\author{Amir Yacoby}
\affiliation{Department of Physics, Faculty of Art and Sciences, Harvard University, Cambridge, MA 02138, USA}

\author{Yuan Cao}
\email{caoyuan@berkeley.edu}
\affiliation{Department of Physics, Faculty of Art and Sciences, Harvard University, Cambridge, MA 02138, USA}
\affiliation{Department of Electrical Engineering and Computer Science, University of California at Berkeley, Berkeley, CA 94720, USA}

\author{Shanhui Fan}
\email{shanhui@stanford.edu}
\affiliation{Department of Applied Physics and Ginzton Laboratory, Stanford University, Stanford, CA 94305, USA}

\author{Eric Mazur}
\email{mazur@seas.harvard.edu}
\affiliation{School of Engineering and Applied Sciences, Harvard University, Cambridge, MA 02138, USA}

\def\thefootnote{$\bot$}\footnotetext{These authors contributed equally to this work}\def\thefootnote{\arabic{footnote}}

\maketitle

\textbf{Reconfigurable optics, optical systems that have a dynamically tunable configuration, are emerging as a new frontier in photonics research. Recently, twisted moiré photonic crystal has become a competitive candidate for implementing reconfigurable optics because of its high degree of tunability. However, despite its great potential as versatile optics components, simultaneous and dynamic modulation of multiple degrees of freedom in twisted moiré photonic crystal has remained out of reach, severely limiting its area of application. In this paper, we present a MEMS-integrated twisted moiré photonic crystal sensor that offers precise control over the interlayer gap and twist angle between two photonic crystal layers, and demonstrate an active twisted moiré photonic crystal-based optical sensor that can simultaneously resolve wavelength and polarization. Leveraging twist- and gap-tuned resonance modes, we achieve high-accuracy spectropolarimetric reconstruction of light using an adaptive sensing algorithm, over a broad operational bandwidth in the telecom range and full Poincaré sphere. Our research showcases the remarkable capabilities of multidimensional control over emergent degrees of freedom in reconfigurable nanophotonics platforms and establishes a scalable pathway towards creating comprehensive flat-optics devices suitable for versatile light manipulation and information processing tasks.}

\begin{figure*}[!ht]
\centering
\includegraphics[width=12cm]{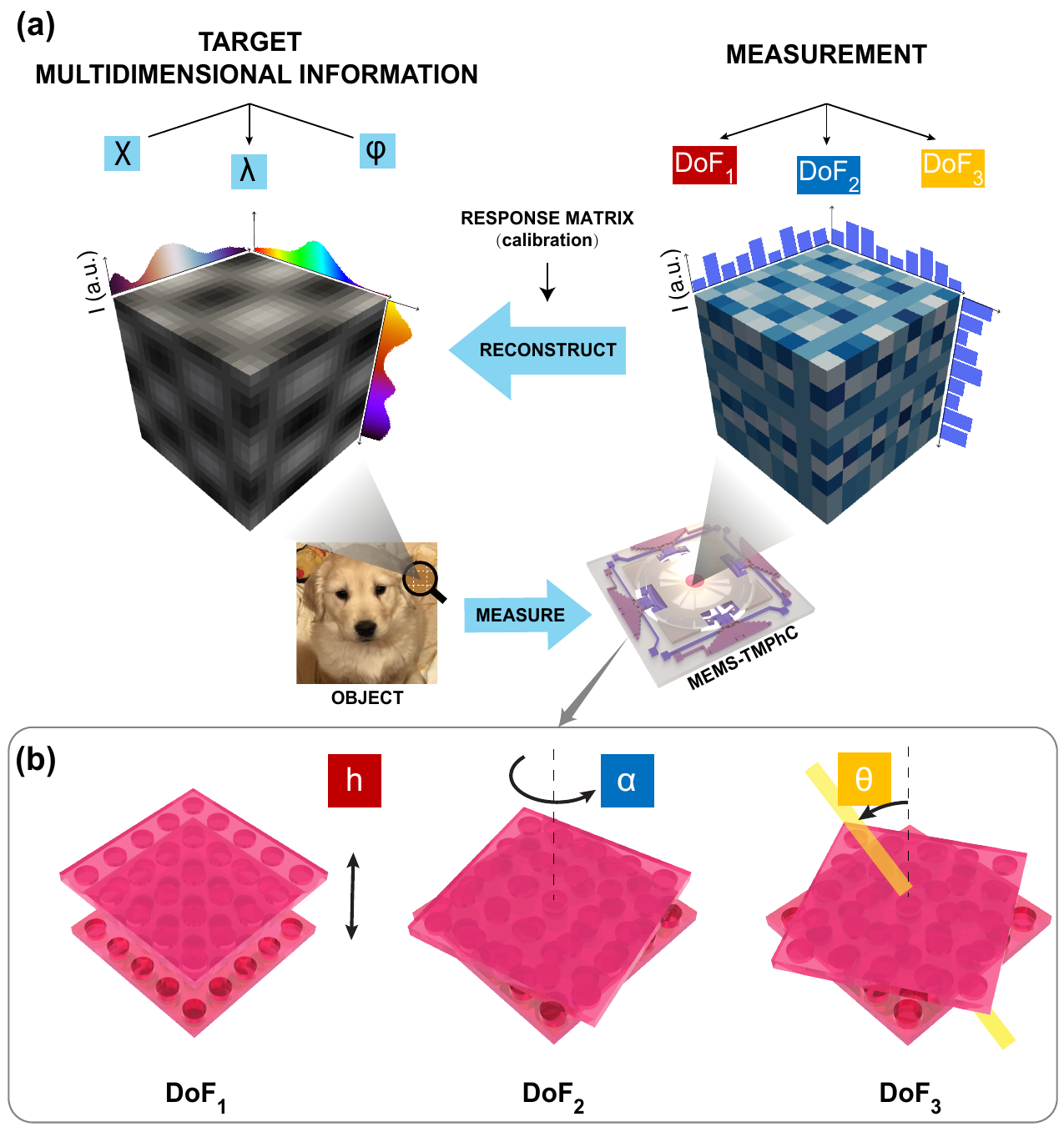}
\caption{\textbf{Computational reconstruction of multidimensional information.} (a) One pixel in an object contains multidimensional information, including wavelength ($\lambda$), ellipticity($\phi$), and azimuth($\chi$). This information can be reconstructed in two steps. First, the intensity of the pixel is measured varying several degrees of freedom ($DoF_1, DoF_2, DoF_3$) using a MEMS-integrated twisted moiré photonic crystal sensor. Next, the target multidimensional information is reconstructed from the measurement matrix. (b) Degrees of freedom for a twisted bilayer moiré photonic crystal include interlayer coupling distance ($h$), twist angle ($\alpha$), and detection angle($\theta$).}
\label{fig:fig1}
\end{figure*}

Extracting phase, polarization, and wavelength information from an optical signal typically necessitates the use of separate optical components, such as interferometers, polarizers, and narrow-band filters or spectrometers, that cannot easily be integrated\cite{yang_miniaturization_2021} and are limited in resolution and speed. Recently, however, computational sensing has made it possible to reconstruct target parameters from the optical signal obtained with a single optical sensor that has multiple configurational degrees of freedom (DoF)\cite{bao_colloidal_2015, yoon_miniaturized_2022, yuan_geometric_2023, wang_single-shot_2019, lee_programmable_2022, meng_detector-only_2020, yang_single-nanowire_2019, yuan_wavelength-scale_2021, kong_single-detector_2021, deng_electrically_2022, ma_intelligent_2022, wang_continuous-spectrumpolarization_nodate, xiong_dynamic_2022,gao_computational_2022,wang_gate-tunable_2020,wang_spectral_2014, ahmed_fully_2021}. In Fig.~\ref{fig:fig1}a, for example, an unknown signal is measured across the three axes of configurational space, where $DoF_1, DoF_2$ and $DoF_3$ represent the configurational DoFs, producing a measurement matrix. After calibrating the sensor by measuring the sensor's response as a function of the three DoFs for each target parameter, the  multidimensional target information can be reconstructed from the measurement and response matrices\cite{yuan_geometric_2023}.  The greater the variation in each response function and the larger the configuration space spanned, the more efficient the reconstruction becomes\cite{jang_wavefront_2018, yu_engineered_2021, gigan_imaging_2022}. In light of these requirements, twisted moiré photonic crystal (TMPhC) — the change in optical properties of a pair of photonic crystals (PhC) as they are twisted relative to each other — stands out as an ideal candidate for computational sensing. 

The optical properties of TMPhC structures result from the strenghtening of the interlayer coupling when two PhCs are placed in close proximity, such that the interlayer spacing $h$ is smaller than the wavelength $\lambda$, which causes electromagnetic waves in each layer to be influenced by those in the other layer. If the two lattices are twisted relative to each other, a moiré superlattice appears in real space. Associated with this superlattice is a new set of wavevectors, leading to what is known as moiré scattering\cite{tang_experimental_2023,lou_theory_2021,lou_tunable_2022,lou_tunable_2022,liu_moire_2022}. In general these moiré photonic structures produce a wide range of tunable optical phenomena in addition to the above mentioned moiré scattering, including
light localization\cite{tang_modeling_2021,tang_-chip_2022,wang_localization_2020,
huang_moire_2022, nguyen_magic_2022, yi_strong_2022,dong_flat_2021,chen_perspective_2021, raun_gan_2023,mao_magic-angle_2021,oudich_photonic_2021}, optical singularities\cite{zhang_twisted_2023,huang_moire_2022,ni_three_2023}, quasiparticle interactions\cite{du_moire_2023,hu_moire_2020,
hu_topological_2020, hu_enhanced_2021, sunku_photonic_2018, chen_configurable_2020,zhang_interface_2021},moiré nonlinearity\cite{krasnok_nonlinear_2018,
fu_optical_2020,arkhipova_observation_2023, yao_enhanced_2021,ha_enhanced_2021,du_twisting_2020,tang_-chip_2023}. The ability to precisely control moiré superlattices is vital for harnessing their unique optical properties and paving the way for future applications such as versatile quantum light sources\cite{fortsch_versatile_2013}, light detection and ranging\cite{park_all-solid-state_2021}, and data communications\cite{salary_time-modulated_2020}. To obtain the multidimensional information required for computational sensing, we exploit moiré scattering from a set of two twisted, identical square-lattice PhCs to modulate the optical response via three DoFs: the interlayer distance ($h$) between the two photonic crystals, their twist angle ($\alpha$), and the detection angle ($\theta$) (Fig.~\ref{fig:fig1}b). These DoFs impact both interlayer and intralayer Bloch state couplings, influencing the moiré scattering. 

While the tunability of TMPhC systems is known in theory, previous experimental studies on twisted moiré structures have typically utilized a fixed device configuration, making them not truly reconfigurable. Recent studies have demonstrated that moiré superlattices with adjustable twist angles can be achieved through mechanical and thermal methods\cite{kapfer_programming_2023,yao_enhanced_2021, wang_thermally_2016,liao_uitra-low_2022,ribeiro-palau_twistable_2018}, which enable one to uncover the intrinsic interlayer-coupling–dependent properties. However, these technologies, apart from being non-integratable, can only modify a single DoF. This constraint limits their effectiveness for consistent TMPhC studies and hinders their use in device applications designed for mass production.

In this paper we experimentally demonstrate the first on-chip dynamically MEMS-integrated twisted moiré photonic crystal (MEMS-TMPhC) device. The device offers dynamic tuning of two DoFs: interlayer spacing $h$ and rotation $\alpha$ (a third DoF, which we do not exploit in this paper, is accessible by varying the detection angle $\theta$) and  permits simultaneous sensing of frequency and polarization information in a single sensor, moving reconfigurable optical experiments from benchtop to chip. Compared to other MEMS-optics devices , our device offers several advantages: it supports multiple DoFs; it features a flat structure that is well-suited for stacking to create more complex devices; and can incorporate additional MEMS-activated DoFs, including lateral translation, tilting, and stretching (see Methods). The fabrication of the sensor utilizes CMOS-compatible wafer-scale processes and can be mass-fabricated using standard foundry nanofabrication processes, ensuring reliability, affordability, and scalability. In addition to computational sensing and imaging, our results offer a pathway towards versatile light manipulation and information processing using multi-DoF MEMS-TMPhC devices.

\section{MEMS-integrated twisted moiré photonic crystal sensor}

\begin{figure*}[!ht]
\centering
\includegraphics[width=\textwidth]{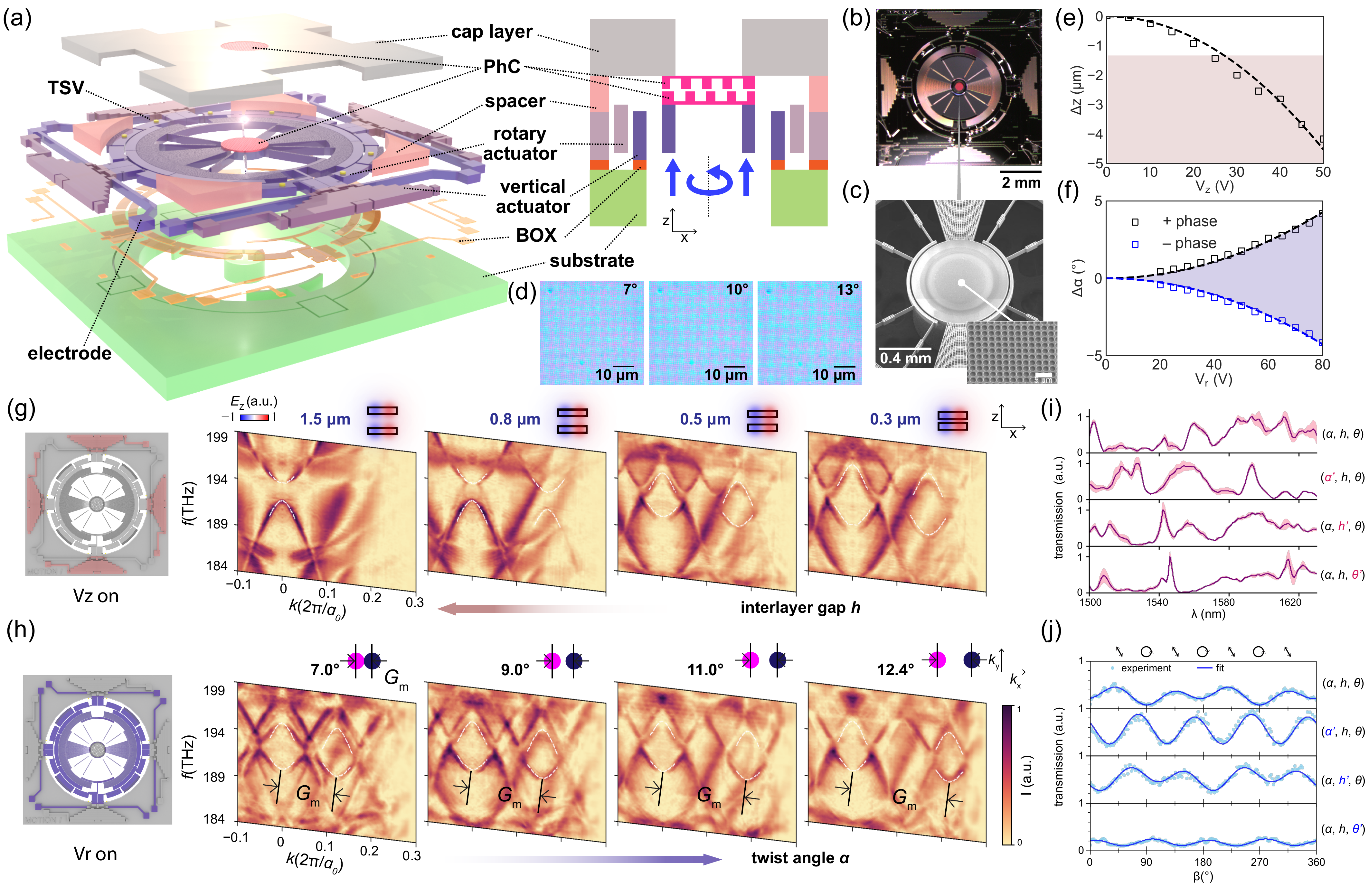}
\caption{\textbf{Microelectromechanically tunable twisted moiré photonic crystal sensor} (a) (right) Schematic of the primary components in a MEMS-TMPhC sensor. PhC = photonic crystal; TSV = through silicon vias; BOX = buffered oxide. (left) Cross-sectional schematic (not to scale) of the MEMS-TMPhC sensor integrating two photonic crystals (PhC) through monolithic fabrication. One of the two PhCs, can be moved vertically and rotationally using MEMS actuators. (b) Microscope image of the MEMS-TMPhC sensor. (c) Scanning electron microscope image of the center of the MEMS-TMPhC sensor, showing one of the integrated PhCs (insert). (d) Microscopic image of the moiré pattern for different rotation angles. (e) Measured and fitted actuation curve for the vertical actuator. The shaded area indicates where a moiré pattern appears. (f) Measured and fitted actuation curve for the rotary actuator. The shaded area indicates the achievable angle range. (g) Left: Vertical actuator. Right: Measured band structures for various vertical spacings $h$. The white dashed lines show the band edges of the corresponding calculated band structures. As the interlayer gap $h$ changes, the band edges shift in frequency. The mode profiles and changing mode coupling is illustrated above each band structure diagram. (h) Left: Rotary actuator. Right: Measured band structures for various twist angles $\alpha$. As the angle grows, the moiré vector lengthens, as illustrated above the band structure diagrams, and the gap $G_m$ between moiré and non-moiré bands expands. (i) Measured transmission spectrum for various values of $\alpha$, $h$, and $k$ (shaded area represents variability in repeated measurements). (j) Transmission for various values of $\alpha$, $h$, and $k$ (shaded area represents variability in repeated measurements). Blue dots are measured data, blue lines are fitted results. Annotations indicate corresponding polarization states for different quarter-wave plate angles.} 
\label{fig:fig2}
\end{figure*}

Figure ~\ref{fig:fig2}a schematically shows the five layers of an MEMS-TMPhC sensor: a \SI{450}{\micro\meter} silicon substrate; a \SI{2}{\micro\meter} buried oxide layer; a \SI{60}{\micro\meter} silicon MEMS-PhC integrated device layer; an SU-8 polymer spacing and bonding layer; and a top cap layer integrated with a fixed PhC. Fig.~\ref{fig:fig2}b provides a top microscope image of the MEMS-TMPhC sensor. The integrated SiN PhC consists of a periodic square lattice of circular holes of \SI{488}{\nano\meter}  radius and \SI{1220}{\nano\meter} spacing (Fig.~\ref{fig:fig2}c). Openings in the layers allow light to be scattered by the moiré PhC. When two PhCs are within \SI{800}{\nano\meter} of each other, a moiré pattern becomes visible under the microscope (Fig.~\ref{fig:fig2}d). See  Methods for the dynamic tuning of the moiré pattern.

The MEMS-integrated PhC can be electrostatically actuated to move vertically or rotate along its axis, using vertical and rotary MEMS actuators that are mechanically connected in series and that are electrically grounded to the silicon handle layer using through-silicon vias (TSV). The vertical actuator is driven by a set of parallel capacitors. Applying V$_z$ to the vertical actuator causes the bottom PhC to move upward  Figure~\ref{fig:fig2}e shows the calculated and measured actuation curves of the vertical MEMS actuator. In the shaded area (V$_z>$\SI{28}{\volt}) the interlayer gap $h<\SI{800}{\nm}$, and a moiré pattern emerges. The rotatory actuator consists of a three-phase electrostatic step motor. When a three-phase alternating voltage is applied, the MEMS-integrated PhC rotates about its center axis. Figure~\ref{fig:fig2}f shows the driving curve for the rotary actuator. When V$_r=$\SI{80}{\volt} the twist angle reaches ±\SI{4.5}{\degree} depending on the phase of the alternating voltage. The shaded area indicate the range of accessible angles at a given applied voltage. See  Methods for the for videos of the two actuators.

Figure.~\ref{fig:fig2}g shows the evolution of the band structure as $h$ changes (corresponding videos of the changing band structure are in Methods). The figure shows both the calculated parabolic band-edges (see Methods) and the band structure measured using a previously published method \cite{tang_experimental_2023}. When $h>\SI{800}{\nm}$, the interlayer coupling is weak, and the band structure is identical to that of a single layer. As $h$ is reduced to $\SI{800}{\nm}$, moiré bands emerge around $0.16k$, introducing secondary parabolic bands near the $\Gamma$-point. A further decrease in $h$ enhances these bands, due to the increased interlayer coupling. Additionally, the frequency of the parabolic band-edge  shifts. Figure.~\ref{fig:fig2}h shows how the band structure is affected by twist angle. The position of the secondary parabolic bands off-$\Gamma$ is influenced by the magnitude of moiré wavevector $\mathcal{G}_{\mathrm{m}}$. As the angle increases from $\SI{7.0}{\degree}$ to $\SI{12.4}{\degree}$, $\mathcal{G}_{\mathrm{m}}$ increases, leading to a shift of the parabolic band-edge from $k=0.24\pi/a$ to $k=0.44\pi/a$, where $a=1220$ nm is the periodicity. The measured band structure is in excellent agreement with both rigorous coupled-wave analysis (RCWA) simulations and the Hamiltonian analytical calculations of the band structure (See Methods). 

To obtain the response matrix, we measure the response functions as a function of  the target variables ($\lambda$, $\chi$,  $\varphi$) across configurational DoF space ($\alpha$, $h$, $\theta$). For example, Fig.~\ref{fig:fig2}i shows the MEMS-TMPhC sensor's response functions as a function of the target variable $\lambda$. The moiré scattering introduces a variety of peaks in the transmission spectrum, and varying the configurational DoFs causes the response function to change significantly. Similarly, to obtain the sensor's response function as a function of polarization state, we measure the sensor's transmission of a fixed-wavelength incident signal varying the polarization of the signal between linear and circular using a linear polarizer and a quarter-wave plate. Figure~\ref{fig:fig2}j shows the sensor's response function as a function of the angle $\beta$ of the quarter-wave plate. The solid curves show a fit of the measured response data to a superposition of circularly and linearly polarized light and a background signal, $T(\beta)=T_{\text {linear }}+T_{\text {circular }}+T_{\text {background }}=T_1 \sin \left(4 \beta+\beta^{\prime\prime}\right)+T_2 \cos \left(2 \beta+\beta^\prime\right)+T_{\text {background }}$, where $T_1$ and $T_2$ determine the configurational parameters $\chi$ and  $\varphi$, respectively\cite{ma_intelligent_2022}. As can be seen, the response curve as a function of polarization state, depends strongly on the configurational DoFs $\alpha$, $h$, and $\theta$, in agreement with previously published results\cite{qin_arbitrarily_2023,overvig_chiral_2021}. 
(See also, Methods)

\section*{Active sensing}

In optical sensing, a common task is to determine the spectral composition and polarization  of light, as represented by $\mathcal{S}(\lambda,p)$, where $S$ denotes the intensity, $\lambda$ denotes the wavelength, $p$ denotes the polarization. In conventional sensing techniques, the intensity distribution $\mathcal{S}(\lambda,p)$ is determined by passing the light through a set of spectral and polarization filters. In computational sensing, one instead first calibrates the device by measuring a set of response functions $\mathcal{R}_\gamma(\lambda, p)$ for a known input signal $\mathcal{S_k}(\lambda, p)$, varying $\gamma$. Next an unknown input signal $\mathcal{S}(\lambda, p)$ is passed through the device, varying $\gamma$, and measures the transmitted intensity $\mathcal{M}_\gamma$, given by
\begin{equation}
    \mathcal{M}_\gamma = \int d\lambda \, dp \; \mathcal{S}(\lambda, p) \, \mathcal{R}_\gamma(\lambda, p).
    \label{eq:meas}
\end{equation}
For our MEMS-TMPhC sensor, $\gamma$ corresponds to a set of values for $(\alpha,h,\theta)$. Finally one uses computational algorithms to extract $S(\lambda,p)$ from the set of responses $\{\mathcal{M}_\gamma\}$. An important aspect in computational sensing is the generation of an appropriate set of response functions $\{ \mathcal{R}_\gamma (\lambda, p)\}$. The set of response functions needs to be sufficiently diverse under different measurement configurations, in order for the signals $\mathcal{S}(\lambda, p)$ to be efficiently reconstructed from the measured responses $\mathcal{M}_\gamma$. The response functions are determined by measuring light transmission under various single wavelengths and polarization states. The wavelength is swept using a narrow-band tunable laser, and the polarization is adjusted by simultaneously rotating a half-wave plate and a quarter-wave plate.

\begin{figure*}[!ht]
\centering
\includegraphics[width=13cm]{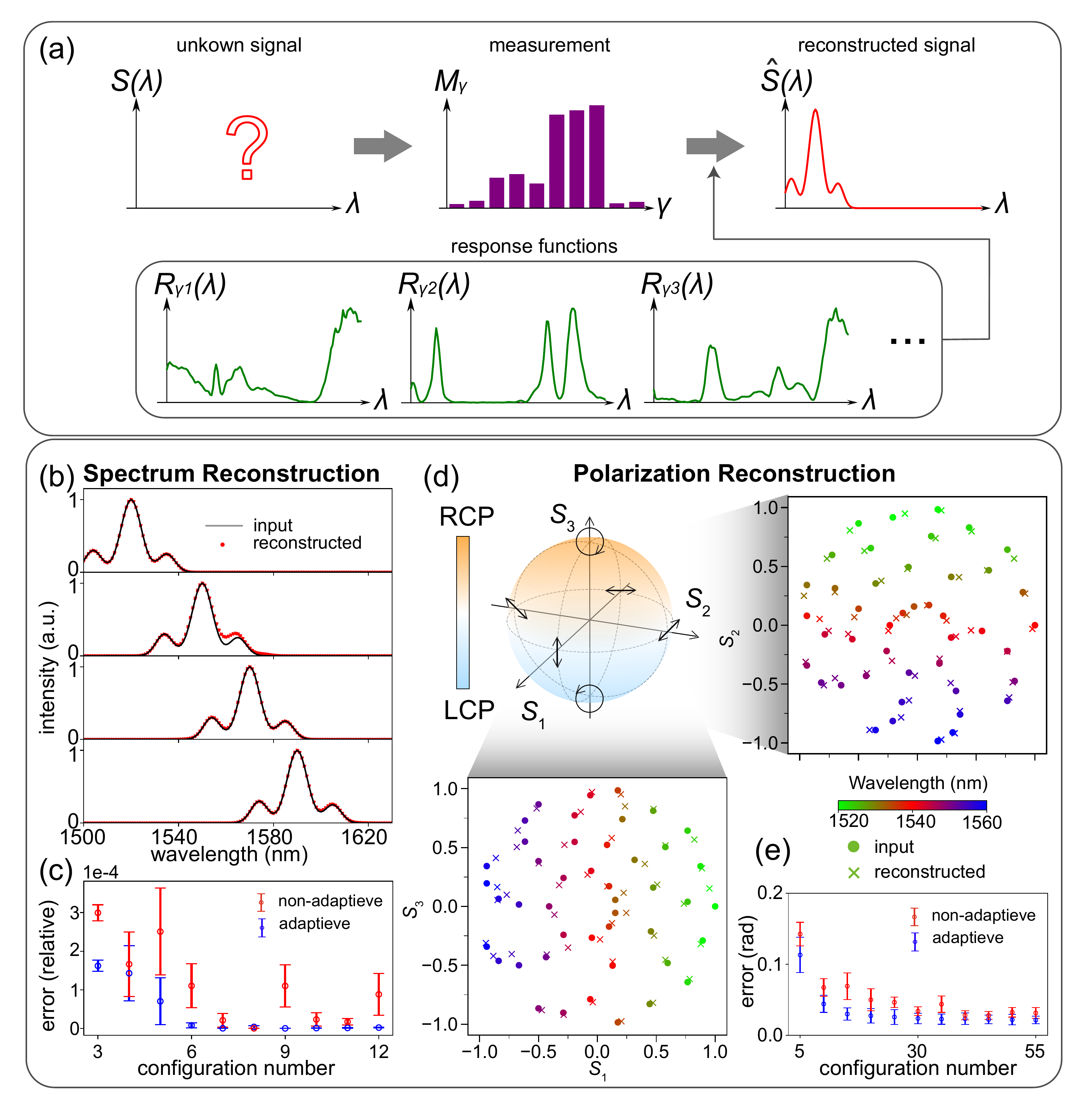}
\caption{\textbf{Sensing principle and results} (a) Input light with unknown frequency composition $\mathcal{S}(\lambda)$ is passed through the sensor, whose response functions $\mathcal{R}_\gamma(\lambda)$ under various configurations $\gamma=\gamma_1,\gamma_2,\gamma_3...$ are known. From the measured intensities $M_\gamma$ under these configurations, one can reconstruct the signal $\hat{\mathcal{S}}(\lambda)$.
(b) Comparison of input and reconstructed spectrum for various signals over the operational frequency range.
(c) Error for spectrum reconstruction as a function of the number of configurations used for intensity measurements, under the adaptive sensing scheme as compared to under the non-adaptive scheme. 
(d) Simultaneous reconstruction of frequency and polarization compositions, where the frequency is represented by color and polarization is represented by location on the Poincaré sphere. The right and bottom panel shows the comparison of input and reconstructed signals, represented by dots and crosses respectively, when both frequency and polarization compositions are unknown.
(e) Error for polarization reconstruction as a function of the number of configurations used for intensity measurements, under the adaptive sensing scheme as compared to under the non-adaptive scheme. 
}
\label{fig:fig3}
\end{figure*}

To illustrate the sensing procedure, we consider an input light signal $\mathcal{S}(\lambda)$ with unknown frequency composition and known polarization, as illustrated in Fig.~\ref{fig:fig3}a.
Under different device configurations $\gamma$, the response functions, as well as the transmitted intensities, are different.
One measures the transmitted intensities under various device configurations and collect these measurement results into a vector $\{ \mathcal{M}_\gamma \}$.
From these measurement results, one can solve an optimization problem to reconstruct the input signal.
Denoting the number of configurations used as $m$ and the measured intensities as $y\in\mathbb{R}^m$, the signal can be reconstructed by solving:
\begin{align}
    \hat{S}(\lambda) &= \argmin_x \,  \lVert Ax-y \rVert_2^2 + \eta_1 \lVert x \rVert_1 
\label{eq:cs}
\end{align}
where $A \in \mathcal{R}^{m \times | \mathcal{B}|}$ is known as the response matrix, with each column corresponding to the expected measurement results for each possible signal under the $m$ configurations used. 
The hyperparameter $\eta_1$ determines the extent of regularization for sparsity, which is commonly used in compressed sensing\cite{eldar_compressed_2012,cs-optical} where the signal is known to be sparse under the chosen set of bases $\mathcal{B}$. 
The solution to Eq.\ref{eq:cs}, denoted as $\hat{S}(\lambda)$, is our reconstructed signal over the set of bases $\mathcal{B}$ (See Methods).

Fig.~\ref{fig:fig3}b shows examples of spectrum reconstruction. The spectra used here are relatively broadband, showing a Gaussian peak with two side bands, which correspond to real spectra commonly encountered in the lab. Examples for other types of spectra, such as narrow-band ones, can be found in Methods. In all the examples, we are able to reconstruct the signal reliably based on the measurements in our sensor.
The dynamic nature of our device allows an adaptive sensing scheme, where one iteratively decides the next measurement configuration $\gamma$ based on previous measurement results. Such adaptive choices of measurement configurations could make each measurement result more informative and therefore make it feasible to sense the signal within very few measurements. 
One common algorithm for adaptive selection of measurement bases is to maximize the information gain based on updated prior knowledge\cite{braun_info-greedy_2014}.
Denoting the sequence of measuremnt configuration choices as $\{ a_1, a_2, ..., a_m\}$ and the corresponding measurement results as $\{ y_1, y_2, ..., y_m \}$, then the each measurement configuration $a_i$ can be chosen based on previous measurement results $\{ y_j\}_{j<i}$:
\begin{equation}
    a_i = \argmax_{a_i\in\mathcal{A}} \mathbb{I}[x; a_i^T x + w_i | \{ y_j, a_j\}_{j<i}]
    \label{eq:adapt}
\end{equation}
where $\mathbb{I}[\cdot;\cdot]$ denotes the mutual information between two random variables, $w_i$ is the measurement noise, and $x$ can be treated as a random variable under Bayesian perspective with the signal distribution treated as the prior probability distribution.
The adaptive scheme is especially useful when the measurement cost is high, such as for specimen that are sensitive to light.

Fig.~\ref{fig:fig3}c shows the advantage of the adaptive sensing scheme, compared to a non-adaptive scheme where the sequence of measurement configurations is simply uniformly sampled in the device configuration space.
Here we show the reconstruction error as a function of number of measurement configurations used, for both adaptive and non-adaptive schemes.
As one increases the number of configurations used, the reconstruction error decreases rapidly, indicating information gain from successive measurements in both schemes.
Under the same number of measurements, the adaptive scheme in general leads to more accurate reconstructions than the non-adaptive one. The advantage of the adaptive scheme is more pronounced when the number of measurements is small. When the number of measurement configurations is large, as the most informative measurement configurations have already been used in the adaptive scheme, the information gain in more measurements for the adaptive scheme will converge to that for the non-adaptive scheme. 
Besides the spectrum reconstruction, we also show the result of full-stoke polarization reconstruction using the same algorithm in Fig.~\ref{fig:fig3}d. Notice that the polarization states and the spectrum information can be identified at the same time. The reconstruction error of adaptive sensing scheme and non-adaptive scheme is shown in Fig.~\ref{fig:fig3}e.

\section*{Hyperspectral and Hyperpolarimetric imaging}

\begin{figure*}[!ht]
\centering
\includegraphics[width=13cm]{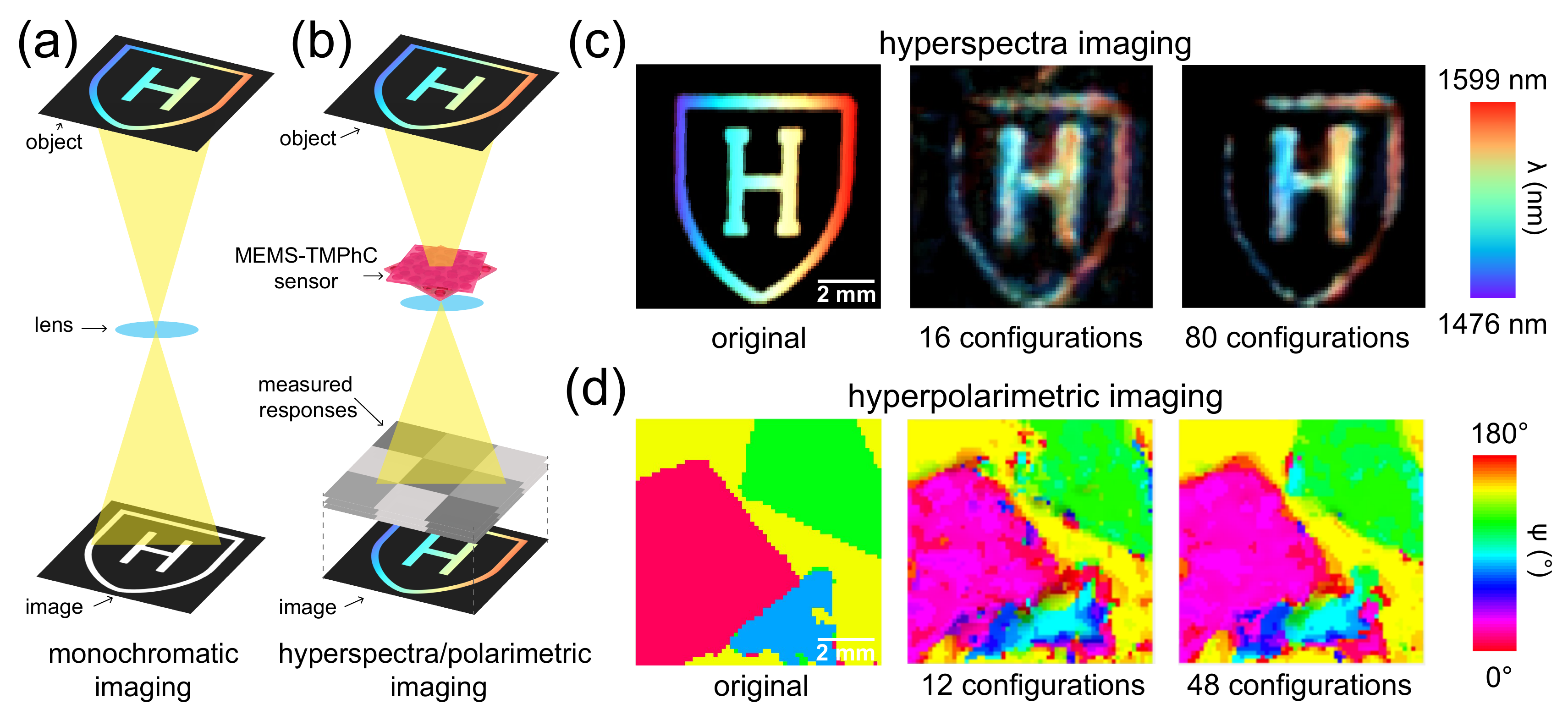}
\caption{\textbf{Hyperspectral and hyperpolarimetric imaging} 
(a) Monochromatic imaging of an object is performed using a monochrome camera. 
(b) Hyperspectral and hyperpolarimetric imaging of an object can be achieved with a monochrome camera when a MEMS-TMPhC sensor is inserted into the optical path, followed by the reconstruction of the measured responses.
(c) compares the object and the hyperspectral imaging results of the Harvard logo under various TMPhC configurations. The spectrum ranges from \SI{1476}{\nano\meter} to \SI{1599}{\nano\meter}
(d) compares the object and the hyperpolarimetric imaging results of three different birefringent materials under various TMPhC configurations. The birefringent materials each have a distinct azimuth angle ($\psi$). 
}

\label{fig:fig4}
\end{figure*}

We demonstrate a proof-of-concept for hyperspectral and hyperpolarimetric imaging using the MEMS-TMPhC sensor.
One unique aspect of our device is that it collects information over a temporal sequence of measurements through a single photonic crystal, rather than using a spatial array of photonic crystals and detectors. As a result, the spatial resolution of our sensor is maximally preserved in imaging tasks.
This feature is highly desirable for hyper-polarimetric and hyper-spectral imaging, which are gaining increasing attention in fields such as remote sensing and medical imaging\cite{yesilkoy_ultrasensitive_2019,faraji-dana_hyperspectral_2019, zuo_chip-integrated_2023, noauthor_matrix_nodate}. In these applications, it is crucial to resolve not only the spectral and polarimetric information but also the spatial distribution of the input signal.
Fig.~\ref{fig:fig4} demostrate both hyperspectral and hyperpolarimetric imaging with the MEMS-TMPhC sensor. 
In conventional monochromatic imaging, the spatial distribution of intensities is measured on the detector of a monochrome camera (Fig.~\ref{fig:fig4}a). In hyperspectral and hyperpolarimetric imaging, a few measurements of the spatial distribution of intensities are taken on the monochrome camera under different TMPhC configurations, allowing for the reconstruction of the signal (Fig.~\ref{fig:fig4}b).
The reconstruction algorithm is the same as previously discussed, except that the signal dimension is higher. Instead of $\mathcal{S}(\lambda,p)$, the signal is $\mathcal{S}(x, y, \lambda)$ in hyperspectral imaging and $\mathcal{S}(x, y, p)$ in hyperpolarimetric imaging. 
In hyperspectral imaging, each pixel of the object (a Harvard logo) exhibits different wavelengths across a continuous spectral range from \SI{1476}{\nano\meter} to \SI{1599}{\nano\meter}. The input signal and the reconstruction under a different number of configurations are shown in Fig.~\ref{fig:fig4}c.
For hyperpolarimetric imaging, the object, containing three pieces of birefringent material, displays three different polarizations ranging from 0° to 180°. The input signal and the reconstruction under different numbers of configurations are shown in Fig.~\ref{fig:fig4}d. The more configurations used, the higher the accuracy that can be obtained for both polarimetric and hyperspectral images.

\section*{Discussion}

We introduced a MEMS-integrated twisted moiré photonic crystal that actively tunes the twist angle and interlayer gap of twisted bilayer photonic crystals. This device enabled us to achieve successful spectrum and polarization reconstruction, as well as proof-of-concept hyperspectral and hyperpolarimetric imaging.
By modulating the twist angle and gap, we observed tunable band structures, wavelengths, and polarizations. The moiré physics in TMPhC structures leads to the creation of a large number of highly tunable resonances in a very compact device.
The sensing capability of our device stems from its diverse and tunable response functions in TMPhC structures. Moreover, the dynamic tuning mechanism enables the implementation of algorithms that utilize prior knowledge or contextual information, allowing for greater flexibility under changing conditions\cite{adaptive-sensing-extreme,adaptive-sensing-astro,she_adaptive_2018}. 
MEMS-TMPhC paves a new pathway for cascaded reconfigurable flat-optics. Our device is fully compatible with on-chip technology and scalability, offering the potential for miniaturization and enhanced tunability. This makes it ideal for a wide range of applications, including versatile sensing, imaging, light manipulation, and lasing.

\end{document}